\documentclass[
reprint,
amsmath,
amssymb,
aps,
]{revtex4-1}

\usepackage{physics}
\usepackage{graphicx}
\usepackage{dcolumn}
\usepackage{bm}
\usepackage{xcolor}
\usepackage{hyperref}
\hypersetup{
	colorlinks=true,       
	linkcolor=blue,        
	citecolor=blue,        
	filecolor=magenta,     
	urlcolor=blue,         
}

\begin{document}


\title{An accelerated linear method for optimizing non-linear wavefunctions in variational Monte Carlo}

\author{Iliya Sabzevari}
\email{iliya.sabzevari@gmail.com}
\affiliation{Department of Chemistry, The University of Colorado at Boulder, Boulder, CO 80302, USA}

\author{Ankit Mahajan}
\affiliation{Department of Chemistry, The University of Colorado at Boulder, Boulder, CO 80302, USA}
 
\author{Sandeep Sharma}
\email{sanshar@gmail.com}
\affiliation{Department of Chemistry, The University of Colorado at Boulder, Boulder, CO 80302, USA}
\homepage{https://www.colorado.edu/lab/sharmagroup/}

\date{\today}

\begin{abstract}
Although the linear method is one of the most robust algorithms for optimizing non-linearly parametrized wavefunctions in variational Monte Carlo, it suffers from a memory bottleneck due to the fact at each optimization step a generalized eigenvalue problem is solved in which the Hamiltonian and overlap matrices are stored in memory. Here we demonstrate that by applying the Jacobi-Davidson algorithm, one can solve the generalized eigenvalue problem iteratively without having to build and store the matrices in question. The resulting direct linear method greatly lowers the cost and improves the scaling of the algorithm with respect to the number of parameters. To further improve the efficiency of optimization for wavefunctions with a large number of parameters, we use the first order method AMSGrad far from the minimum as it is very inexpensive, and only switch to the direct linear method near the end of the optimization where methods such as AMSGrad have long convergence tails. We apply this improved optimizer to various wavefunctions with both real and orbital space Jastrow factors for atomic systems such as Beryllium and Neon, molecular systems such as the Carbon dimer and Iron(II) Porphyrin, and model systems such as the Hubbard model and Hydrogen chains.
\end{abstract}

\maketitle


\section{\label{sec:1} Introduction}

Many wavefunction based electronic structure methods rely upon the variational principle, which states that the expectation value of the Hamiltonian for a given wavefunction is an upper bound to the true ground state energy of the Hamiltonian. Thus, searching for an approximate ground state solution amounts to minimizing the energy of an appropriately chosen ansatz. Given such an ansatz $\ket{\Psi(\mathbf{p})}$, we seek the stationary point of the energy functional $E = \frac{\bra{\Psi(\mathbf{p})}\hat{H}\ket{\Psi(\mathbf{p})}}{\bra{\Psi(\mathbf{p})}\ket{\Psi(\mathbf{p})}}$, with respect to the parameters $\{\mathbf{p}_i\}$. 
If the wavefunction ansatz is a linear function of the parameters, the resulting optimization problem becomes relatively easy and many robust algorithms are available. However, in variational Monte Carlo (VMC)\cite{corrsys,realspace} the most commonly used wavefunctions depend non-linearly on the parameters. Although this non-linearity leads to great flexibility in the wavefunction, it also introduces strong coupling between the parameters and results in a very challenging non-linear optimization problem. In addition, in Monte Carlo algorithms, all quantities such as energies, gradients and Hessians are only known up to a stochastic error which renders several commonly used algorithms for deterministic optimization ineffective.

Only in the past decade have approaches been developed within VMC that are able to optimize the energy of a wavefunction with more than a few thousand parameters. These techniques include the stochastic reconfiguration (SR) algorithm\cite{sorella2001generalized,sorella2005wave,beaudet2008molecular} and its low-memory extension introduced by Neuscamman \textit{et al.}\cite{neuscamman2012optimizing}, and adaptive stochastic gradient descent (SGD) algorithms developed in the machine learning community utilized by Booth and coworkers\cite{schwarz2017projector} and our group\cite{sabzevari2018improved}. This has led the way to optimization of wavefunction ansatzes with up to $\sim10^5$ parameters. The above algorithms are termed first order optimization methods, as they only rely upon information from first derivatives.

Another very promising approach that has worked extremely well in the face of these challenges is the linear method (LM), which distinguishes itself from the former algorithms by using some information from second derivatives\cite{nightingale2001optimization,umrigar2007alleviation,toulouse2007optimization,toulouse2008full}. This algorithm amounts to solving a generalized eigenvalue problem in the self plus tangent space of the current wavefunction. The LM utilizes a strong \textit{zero-variance} property, and is the method of choice for fast and robust optimization for relatively small systems\cite{neuscamman2016improved,goetz2017suppressing,brown2007energies,petruzielo2012approaching}. It has been a challenge to apply the LM to larger systems due to the substantial memory requirement of storing the matrices and the large expense of solving the generalized eigenvalue problem. Neuscamman and coworkers have recently introduced a variant of the LM called the Blocked Linear Method (BLM)\cite{zhao2017blocked} that has been applied to wavefunctions containing up to $\sim25,000$ parameters. In this work we follow an alternative route to extending the LM for large systems. We do so by by employing Davidson's method with the Jacobi orthogonal component correction to solve the generalized eigenvalue problem iteratively. This yields a solution that only requires repeated action of the Hamiltonian and overlap matrices onto a vector, which can be done at a low scaling cost as discussed by Neuscamman \textit{et al.}\cite{neuscamman2012optimizing}. In addition, by sampling the matrices after a correlation length of moves have been made, one can reduce the space and time complexity of applying these matrices. Further, to tackle truly large systems, we also use AMSGrad, a significantly cheaper optimizer, far from the minimum where the first order optimization methods work very well\cite{reddi2019convergence} and only switch over to the direct linear method near convergence. These ideas allow us to apply this accelerated LM (aLM) to wavefunctions with up to $\sim60,000$ parameters, with promising outlook for $10^5$ parameters.

The rest of this article is organized as follows. In section \ref{sec:2} we outline the theory and methods employed, beginning with the VMC framework our group has developed in real and orbital space, and then moving onto the theory of the aLM. Section \ref{sec:3} goes over the finer computational details of the overall algorithm, and in section \ref{sec:4} we show numerical experiments for various systems from chemistry and physics. We end with our final remarks in section \ref{sec:5}, the conclusion, and \ref{sec:6}, acknowledgments.

\section{\label{sec:2}Theory and Methods}

\subsection{Variational Monte Carlo Framework}

In the past year, we have developed an improved orbital space variational Monte Carlo (VMC) framework in which the overall cost of the algorithm for obtaining a system size independent stochastic error was lowered from $O(N^5)$ to $O(N^4)$\cite{sabzevari2018improved} (Neuscamman and Wei have achieved the same reduction in cost with a semi-stochastic algorithm\cite{wei2018reduced}). Here $N$ is a general measure of system size, and is of the same order as the number of orbitals and the number of electrons. The key improvements include:

\begin{enumerate}
    \item The use of an effective Hamiltonian that reduces the cost of the local energy calculation from $O(N^4)$ to $O(N^2)$; inspired by the Heat-bath Configuration Interaction (HCI) algorithm\cite{holmes2016heat}.
    \item The use of continuous time Monte Carlo for sampling occupation number vectors that ensures ergodic and efficient moves\cite{bortz1975new,gillespie1976general}.
    \item The use of AMSGrad to optimize parameters\cite{reddi2019convergence}. This is a stochastic gradient descent algorithm developed in the machine learning community that yields comparable convergence to the popular stochastic reconfiguration (SR) algorithm\cite{neuscamman2012optimizing,sabzevari2018improved}.
\end{enumerate}

We have recently published an article expanding this framework with highly accurate wavefunction ansatzes consisting of a Jastrow factor and a projected symmetry broken mean-field reference\cite{mahajan2019symmetry}. These wavefunctions have the form
\begin{equation}
    \ket{\Psi} = \hat{J}\hat{P}\ket{\Phi},
\end{equation}
where $\ket{\Phi}$ is a reference ansatz with greater variational freedom that breaks certain symmetries, $\hat{P}$ is a symmetry projector that projects out the parts of the reference that satisfy the symmetries of the Hamiltonian, and $\hat{J}$ is a Jastrow operator that adds explicit correlation between electrons. It has been demonstrated that a variational ansatz of the form $\hat{P}\ket{\Phi}$ has more flexibility than an ansatz $\ket{\Phi}$ that obeys the symmetries of the Hamiltonian by construction\cite{bach1994generalized,lykos1963discussion,lowdin1955quantum}. We follow prior work done in the community, and project the correct symmetries before varying the parameters\cite{scuseria2011projected,rodriguez2012symmetry,jimenez2012projected}. For this paper, the symmerty broken reference will be taken to be a generalized Hartree Fock (GHF) ansatz.
GHF wavefunctions are a product of fermionic creation operators acting on the vacuum state,
\begin{equation}
    \ket{\text{GHF}} = \prod_{k=1}^{n_{\text{elec}}}\hat{a}_k^{\dagger}\ket{0},
\end{equation}
with the creation operators defined as linear combinations of local atomic orbital creation operators,
\begin{equation}
    \hat{a}_k^{\dagger} = \sum_{l=1}^{n_{\text{orb}}}\sum_{\sigma=\uparrow,\downarrow}\theta_{l\sigma,k} \cdot 
    \hat{c}_{l\sigma}^{\dagger}.
\end{equation}
Note that the sum is over all spin projections, hence the GHF wavefunction breaks both $\hat{S}^2$ and $\hat{S}_z$ symmetries. By using complex variables for $\theta$, complex conjugation ($\hat{K}$) symmetry can also be broken. We then restore $\hat{K}$ and $\hat{S}_z$ symmetries with the aforementioned projector. The orbital space Jastrow factor has the form
\begin{equation}
    \hat{J} = \exp{\sum_{p \geq q}^{n_{orb}} \sum_{\sigma,\gamma=\uparrow,\downarrow} J_{p\sigma,q\gamma}\hat{n}_{p\sigma}\hat{n}_{q\gamma}},
\end{equation}
where $\hat{n}_{p\sigma}$ and $\hat{n}_{q\gamma}$ are number operators for the spin orbitals $p\sigma$ and $q\gamma$, respectively, and $J_{p\sigma,q\gamma}$ are variational parameters. A Jastrow factor such as this is part of a larger class of correlators which include correlated product states\cite{mezzacapo2009ground,neuscamman2012correlator}, and Gutzwiller factors\cite{gutzwiller1963effect}. This form ensures that the full ansatz is size consistent by using local number projectors\cite{neuscamman2012size,neuscamman2013jastrow} and also imparts some dynamical correlation\cite{neuscamman2013communication,neuscamman2016subtractive}. The inclusion of a linearized Jastrow-type factor with a spin symmetry projected GHF wavefunction has been used before by Scuseria and Henderson\cite{henderson2013linearized}. In the context of VMC, many studies have utilized an ansatz such as this for model Hamiltonians\cite{tahara2008variational,tahara2008diff,kurita2015variational,zhao2017variational,darmawan2018stripe,misawa2019mvmc}.

In addition to the above work, we have recently extended symmetry broken references to real space VMC. The GHF ansatz has the same expression, however the Jastrow factor has a different form in this basis,
\begin{equation}
    \hat{J} = \exp{\sum_{i,I}f(r_{iI}) + \sum_{i>j}g(r_{ij}) + \sum_{i,j,I} h(r_{ij}, r_{iI}, r_{jI})},
\end{equation}
where $r_{ij}$ is the distance between electrons $i$ and $j$, and $r_{iI}$ is the distance between electron $i$ and nucleus $I$. The first summation represents the one body electron-nucleus correlations, the second summation represents the two body electron-electron correlations, and the third summation represents the three body electron-electron-nucleus correlations. The functions $f$, $g$, and $h$ are usually of some polynomial form, here we take them to be the polynomials introduced by Boys and Handy\cite{boys1969calculation}. This Jastrow factor is chosen to satisfy the electron-electron and electron-nuclear cusp conditions of the true many-body wavefunction, which account for much of the shortcomings of the approximate wavefunctions used throughout quantum chemistry\cite{kato1957eigenfunctions,drummond2004jastrow,purple}.
Details of the real space algorithm and wavefunctions will be discussed in a forthcoming publication. For more information on the orbital space algorithm, please refer to our previous publications\cite{mahajan2019symmetry,sabzevari2018improved}.

\subsection{The Linear Method}

The variational parameters defining the wavefunction ansatz need to be optimized in accordance with the variational principle; we accomplish this with the accelerated Linear Method (aLM). Here we outline our approach, following previous work done in the community\cite{toulouse2007optimization,toulouse2008full,umrigar2007alleviation,nightingale2001optimization,neuscamman2012optimizing}. The LM begins by defining the following $n_{\text{var}} + 1$ dimensional basis at any point in parameter space for a given normalized ansatz $\ket{\Psi}$.

\begin{equation}\label{eq:1}
    \ket{\bar{\Psi}_i} = 
    \begin{cases}
        \ket{\Psi}, & \text{if} \: i = 0 \\
        \ket{\Psi_i} - \ket{\Psi} \bra{\Psi}\ket{\Psi_i} , & \text{if} \: i \neq 0 \\
    \end{cases}
\end{equation}
where $\ket{\Psi_i} = \frac{\partial}{\partial p_i}\ket{\Psi}$. The wavefunction can then be expanded as $\ket{\Psi_{\text{lin}}} = \sum_{i} \ket{\bar{\Psi}_i}\delta p_i$, with $\delta p_0 = 1$. By minimizing the energy of this linearized wavefunction with respect to the parameters $\{\delta p_i\}$, one obtains the following generalized eigenvalue problem:
\begin{equation} \label{eq:gep}
    \mathbf{\bar{H}} \cdot \delta\mathbf{p} = E \cdot \mathbf{\bar{S}} \cdot \delta\mathbf{p},
\end{equation}
where we have defined $\mathbf{\bar{H}}_{ij} = \bra{\bar{\Psi}_i} \hat{H} \ket{\bar{\Psi}_j}$ and $\mathbf{\bar{S}}_{ij} = \bra{\bar{\Psi}_i}\ket{\bar{\Psi}_j}$. This eigenvalue problem can be solved to find the ideal parameter updates $\{\delta p_i\}$. However, there is an arbitrariness in this procedure, as the normalization constant of $\ket{\Psi}$ is a function of the parameters and how it is defined can lead to unwieldy parameter updates. To take into account this ambiguity, we follow Umrigar and Toulouse and perform the following uniform scaling on the parameter updates\cite{toulouse2008full,toulouse2007optimization}:

\begin{equation}
    \delta p_i' = \frac{\delta p_i}{1 - \sum_{i = 1}^{n_{var}} N_i \cdot \delta p_i},
\end{equation}
where for linear parameters,

\begin{equation}
    N_i = \bra{\Psi}\ket{\Psi_i},
\end{equation}
and for non-linear parameters,
\begin{equation}
    N_i = -\frac{(1-\xi) \cdot \sum_{j}^{nonlin} \mathbf{\bar{S}}_{ij} \cdot \delta p_j}{(1-\xi) + \xi \cdot \sqrt{1 + \sum_{ij}^{nonlin} \delta p_i \cdot \mathbf{\bar{S}}_{ij} \cdot \delta p_j}}.
\end{equation}
In this work, $\xi = 0.5$ for all calculations.

The matrices, $\mathbf{\bar{H}}$ and $\mathbf{\bar{S}}$ can be calculated via Monte Carlo sampling by inserting the resolution of identity.
\begin{align*}
    \mathbf{\bar{H}}_{ij} &= \sum_n \bra{\bar{\Psi}_i}\ket{n} \bra{n}\hat{H}\ket{\bar{\Psi}_j} \\
    &= \sum_n \frac{\bra{\bar{\Psi}_i}\ket{n}}{\bra{\Psi}\ket{n}} \cdot \frac{\bra{n}\hat{H}\ket{\bar{\Psi}_j}}{\bra{n}\ket{\Psi}} \cdot \lvert\bra{n}\ket{\Psi}\rvert^2
\end{align*}
\begin{equation}
    \mathbf{\bar{H}}_{ij} = \expval{\frac{\bra{\bar{\Psi}_i}\ket{n}}{\bra{\Psi}\ket{n}} \cdot \frac{\bra{n}\hat{H}\ket{\bar{\Psi}_j}}{\bra{n}\ket{\Psi}}}_{\rho_n},
\end{equation}
where we have defined a probability distribution $\rho_n = \lvert\bra{n}\ket{\Psi}\rvert^2$. It is important to note the matrix $\mathbf{\bar{H}}$ will be symmetric when the complete sum is performed, however, this will not be the case when a Monte Carlo average is calculated\cite{nightingale2001optimization}. A similar result is obtained with respect to the overlap matrix,
\begin{equation}
    \mathbf{\bar{S}}_{ij} = \expval{\frac{\bra{\bar{\Psi}_i}\ket{n}}{\bra{\Psi}\ket{n}} \cdot \frac{\bra{n}\ket{\bar{\Psi}_j}}{\bra{n}\ket{\Psi}}}_{\rho_n}.
\end{equation}
Inserting explicit expressions for the basis in Eq.(\ref{eq:1}), it is evident that $\mathbf{\bar{S}}_{00} = 1$ and $\mathbf{\bar{H}}_{00} = \expval{E_L[n]} = E_0$.
Where we recognize a local energy, $E_L[n] = \frac{\bra{n}\hat{H}\ket{\Psi}}{\bra{n}\ket{\Psi}}$, which when averaged is the energy of the zeroth order wavefunction $\ket{\Psi}$. It is convenient to introduce the quantities $\mathbf{g}_i[n]$, and $\mathbf{h}_i[n]$ which are defined as:
\begin{equation}
    \mathbf{g}_i[n] = \frac{\bra{n}\ket{\Psi_i}}{\bra{n}\ket{\Psi}},
\end{equation}
\begin{equation}
    \mathbf{h}_i[n] = \frac{\bra{n}\hat{H}\ket{\Psi_i}}{\bra{n}\ket{\Psi}}.
\end{equation}
We can then write the matrix equation as:
\begin{equation}
    \begin{pmatrix}
        E_0 & \mathbf{G}_r^T \\
        \mathbf{G}_c & \mathbf{H}
    \end{pmatrix}
    \begin{pmatrix}
        1 \\
        \delta \mathbf{p}
    \end{pmatrix}
    =
    E
    \begin{pmatrix}
        1 & \mathbf{0}^T \\
        \mathbf{0} & \mathbf{S}
    \end{pmatrix}
    \begin{pmatrix}
        1 \\
        \delta \mathbf{p}
    \end{pmatrix},
\end{equation} \\
where $\mathbf{G_r}$ and $\mathbf{G_c}$ are two different estimates of the energy gradient (neglecting a factor of $2$), with dimension $n_{\text{var}}$,
\begin{equation}
   \mathbf{G_r}_{i} = \expval{\mathbf{h}_i[n]} - E_0 \cdot \expval{\mathbf{g}_i[n]}
\end{equation}
and
\begin{equation}
    \mathbf{G_c}_{i} = \expval{E_L[n] \cdot \mathbf{g}_i[n]} - E_0 \cdot \expval{\mathbf{g}_i[n]}.
\end{equation}
The newly introduced matrices $\mathbf{H}$ and $\mathbf{S}$ are $n_{\text{var}} \cdot n_{\text{var}}$ dimensional matrices with the form,
\begin{equation}
    \mathbf{S}_{ij} = \expval{\mathbf{g}_i[n] \cdot \mathbf{g}_j[n]} - \expval{\mathbf{g}_i[n]} \cdot \expval{\mathbf{g}_j[n]},
\end{equation}
\begin{equation}
    \begin{aligned}
        \mathbf{H}_{ij} = \expval{\mathbf{g}_i[n] \cdot \mathbf{h}_j[n]} - \expval{\mathbf{g}_i[n] \cdot E_L[n]} \cdot \expval{\mathbf{g}_j[n]} \\ -
        \expval{\mathbf{g}_i[n]} \cdot \expval{\mathbf{h}_j[n]} + \expval{\mathbf{g}_i[n]} \cdot E_o \cdot \expval{\mathbf{g}_j[n]}.
    \end{aligned}
\end{equation}

This rewriting of Eq.(\ref{eq:gep}), together with the introduction of
\begin{equation}
    \mathbf{\bar{g}}_i[n] =
    \begin{cases}
        0, \text{ for } i = 0 \\
        \mathbf{g}_i[n], \text{ for }  i \neq 0
    \end{cases},
\end{equation}
\begin{equation}
    \mathbf{\bar{h}}_i[n] =
    \begin{cases}
        0, \text{ for } i = 0 \\
        \mathbf{h}_i[n], \text{ for }  i \neq 0
    \end{cases},
\end{equation}
and of
\begin{equation}
    \mathbf{\bar{G}_r}{}_i =
    \begin{cases}
        0, \text{ for } i = 0 \\
        \mathbf{G_r}_{i}, \text{ for }  i \neq 0
    \end{cases},
\end{equation}
\begin{equation}
    \mathbf{\bar{G}_c}{}_i =
    \begin{cases}
        0, \text{ for } i = 0 \\
        \mathbf{G_c}_{i}, \text{ for }  i \neq 0
    \end{cases},
\end{equation}
allow the matrices $\mathbf{\bar{H}}$ and $\mathbf{\bar{S}}$ to be compactly written as a sum of outer products of vectors,
\begin{equation}
    \mathbf{\bar{S}} = \mathbf{e_0} \cdot \mathbf{e_0}^T + \expval{\mathbf{\bar{g}} \cdot \mathbf{\bar{g}}^T} - \expval{\mathbf{\bar{g}}} \cdot \expval{\mathbf{\bar{g}}}^T,
\end{equation}
\begin{equation}
    \begin{aligned}
        \mathbf{\bar{H}} = E_0 \cdot \mathbf{e_0} \cdot \mathbf{e_0}^T + \mathbf{e_0} \cdot \mathbf{\bar{G}_r}^T + \mathbf{\bar{G}_c} \cdot \mathbf{e_0}^T \\
        + \expval{\mathbf{\bar{g}} \cdot \mathbf{\bar{h}}^T} - \expval{\mathbf{\bar{g}} \cdot E_L} \cdot \expval{\mathbf{\bar{g}}}^T \\
        - \expval{\mathbf{\bar{g}}} \cdot \expval{\mathbf{\bar{h}}}^T + \expval{\mathbf{\bar{g}}} \cdot E_0 \cdot \expval{\mathbf{\bar{g}}}^T.
    \end{aligned}
\end{equation}
Here $\mathbf{e_0}$ is the orthogonal unit vector of index $0$ and dimension $n_{\text{var}} + 1$. This form suggests that, instead of building these matrices at a cost of $O(n_{\text{s}}  (n_{\text{var}})^2)$ (where $n_{\text{s}}$ is the number of samples), the action of the matrices on a vector $\mathbf{z}$ can be calculated at a cost of $O(n_{\text{s}} n_{\text{var}})$:
\begin{equation}
    \mathbf{\bar{S}} \cdot \mathbf{z} = \mathbf{e_0} \cdot (\mathbf{e_0}^T \cdot \mathbf{z}) + \expval{\mathbf{\bar{g}} \cdot (\mathbf{\bar{g}}^T \cdot \mathbf{z})} - \expval{\mathbf{\bar{g}}} \cdot (\expval{\mathbf{\bar{g}}}^T \cdot \mathbf{z}),
\end{equation}
and likewise with the $\mathbf{\bar{H}}$ matrix.
This amounts to a series of dot products, scalar multiplications, and addition of vectors. Although this improves the time complexity of the action of the matrices, it introduces a large memory cost by requiring the storage of the primitive values $E_L[n]$, $\mathbf{g}_i[n]$,
and 
$\mathbf{h}_i[n]$
for the entire Monte Carlo simulation over the sampled states $\{n\}$. While this seems excessive, in practice this seemingly large memory cost is still usually smaller than that of the non-direct algorithm in which the matrices $\mathbf{\bar{H}}$ and $\mathbf{\bar{S}}$ are stored in memory. This is because as the VMC algorithm is parallelized across $n_{\text{proc}}$ processors, the memory requirement per processor for the direct algorithm reduces to $O(n_{\text{s}} n_{\text{var}}/ n_{\text{proc}})$, while the memory requirement per processor for the non-direct linear method remains $O((n_{\text{var}})^2)$. 
As a result, as the degree of parallelization increases, the memory requirement per process for the direct algorithm improves relative to the non direct algorithm. 
Furthermore, the matrices can be sampled after a correlation length of moves have been made. This leads to a $n_{\text{corr}}$ factor reduction in the space complexity of storing the primitives and the time complexity of applying the matrices onto a vector, yielding a $O(n_{\text{s}} n_{\text{var}} / n_{\text{corr}}/n_{\text{proc}})$ asymptotic CPU and memory cost per process.

\subsection{Davidson's method and the Jacobi orthogonal component correction}
In this subsection, we will discuss how to solve the introduced generalized eigenvalue problem with only access to the action of the matrices onto a vector. In this formulation of the LM, we can make use of iterative solvers such as Davidson's method\cite{davdison1975iterative}, which allow one to calculate a single or a few close together eigenvalues and eigenvectors without explicit construction of the matrices\cite{davdison1975iterative,morgan1986generalizations,morgan1992generalizations}. We will only focus on calculating a single eigenpair $(\lambda, \mathbf{v})$ for the generalized eigenproblem,
\begin{equation}
    \mathbf{A} \cdot \mathbf{v} = \lambda \cdot \mathbf{B} \cdot \mathbf{v}.
\end{equation}
$\mathbf{A}$ is a general square matrix and $\mathbf{B}$ is the overlap matrix or \textit{metric}, which is positive semi-definite. Assuming we have a reasonable guess $\mathbf{x}$ to the eigenvector $\mathbf{v}$, we seek a correction $\mathbf{\delta}$,
\begin{equation}
    \mathbf{A} \cdot (\mathbf{x} + \mathbf{\delta}) = \rho \cdot \mathbf{B} \cdot (\mathbf{x} + \mathbf{\delta}),
\end{equation}
where $\lambda$ is replaced by the Rayleigh quotient of our guess $\rho = \frac{\mathbf{x}^\dagger \mathbf{A} \mathbf{x}}{\mathbf{x}^\dagger \mathbf{B} \mathbf{x}}$. Rearranging the above equation yields,
\begin{equation} \label{eq:correction}
    (\mathbf{A} - \rho \cdot \mathbf{B}) \cdot \mathbf{\delta} = - \mathbf{r},
\end{equation}
where $\mathbf{r}$ is the so called residual vector equal to $(\mathbf{A} - \rho \cdot \mathbf{B}) \cdot \mathbf{x}$. This can conceivably be solved for the correction, but in reality gives a poor update due to the fact $\mathbf{x}$ and $\mathbf{\delta}$ can be linearly dependent. In practice, an approximate correction equation is solved,
\begin{equation} \label{eq:correction1}
    \mathbf{F} \cdot \mathbf{\delta} = - \mathbf{r},
\end{equation}
where $\mathbf{F}$ is a \textit{preconditioner}, ie. some matrix similar to $(\mathbf{A} - \rho \cdot \mathbf{B})$. Davidson's original paper was concerned with problems where $\mathbf{B} = \mathbf{I}$, and by assuming $\mathbf{A}$ is well approximated by it's diagonal values, he came to the following expression for the correction,
\begin{equation} \label{eq:perturb}
    \mathbf{\delta} = \frac{-\mathbf{r}}{\mathbf{A_{diag}} - \rho \cdot \mathbf{I}},
\end{equation}
where $\mathbf{A_{diag}}$ is the matrix containing only the diagonal of $\mathbf{A}$. Eq.(\ref{eq:perturb}) can be recognized as a perturbative correction to $\mathbf{x}$ and is very easy to calculate. For diagonally dominant matrices like many of those in quantum chemistry, this correction is remarkably successful\cite{davdison1975iterative,van1996improvement,purple}. The $\mathbf{\bar{H}}$ matrix is not diagonally dominant and the $\mathbf{\bar{S}}$ matrix is not equal to the identity, due to the aforementioned strong coupling of the wavefunction parameters. This forces us to find other expressions for the matrix $\mathbf{F}$.

The Jacobi orthogonal component correction (JOCC) attempts to find a correction $\mathbf{\delta}$, that is orthogonal to the current best guess $\mathbf{x}$. It does this by solving a projected form of Eq.(\ref{eq:correction}) with
\begin{equation}
    \mathbf{F} = \mathbf{P}^\dagger (\mathbf{A} - \rho \cdot \mathbf{B})\mathbf{P}.
\end{equation}
Where $\mathbf{P} = (\mathbf{I} - \mathbf{x} \cdot \mathbf{y}^\dagger)$ and $\mathbf{y} = \mathbf{B} \cdot \mathbf{x}$\cite{sleijpen1996jacobi}. This  $\mathbf{F}$ matrix happens to be positive semi-definite (with the addition of a diagonal shift to $\mathbf{A}$) and only requires a handful of iterations of conjugate gradient to generate a reasonable correction. Alternatively, one can also set a loose convergence threshold for $\mathbf{\delta}$ compared to the tolerance required of $\mathbf{x}$. Davidson's method with the JOCC is termed the Jacobi-Davidson method, and generalizes Davidson's method to \textit{any} matrix, no matter the magnitude of the off-diagonal elements\cite{sleijpen1996jacobi,sleijpen2000jacobi,bai2000templates}.

Here we'll outline the algorithmic details of Davidson's method. The goal is to solve the eigenvalue problem in a gradually growing subspace of orthonormal vectors with the hope that the solution exists in the span of this subspace. A general Davidson's algorithm with input matrices $\mathbf{A}$ and $\mathbf{B}$, a target value $t$, and a reasonable guess vector $\mathbf{x_0}$, has the following series of steps.
\begin{enumerate}
    \item Initialize the subspace matrices $\mathbf{V} = [\mathbf{x_0}]$, $\mathbf{V_A} = [\mathbf{A}\mathbf{x_0}]$, and $\mathbf{V_B} = [\mathbf{B}\mathbf{x_0}]$.
    \item Solve the subspace problem,
    \begin{equation}
        \mathbf{A^\prime} \cdot \mathbf{s} = \theta \cdot \mathbf{B^\prime} \cdot \mathbf{s},
    \end{equation}
    where $\mathbf{A^\prime} = \mathbf{V}^\dagger \mathbf{A} \mathbf{V} = \mathbf{V}^\dagger \mathbf{V_A}$ and $\mathbf{B^\prime} = \mathbf{V}^\dagger \mathbf{B} \mathbf{V} = \mathbf{V}^\dagger \mathbf{V_B}$. Select the eigenpair $(\theta, \mathbf{s})$ in which $\theta$ is closest to $t$.
    \item Transform $\mathbf{s}$ into the original problem space, $\mathbf{u} = \mathbf{V} \cdot \mathbf{s}$, and calculate action of matrices, $\mathbf{u_A} = \mathbf{A} \cdot \mathbf{u} = \mathbf{V_A} \cdot \mathbf{s}$ and $\mathbf{u_B} = \mathbf{B} \cdot \mathbf{u} = \mathbf{V_B} \cdot \mathbf{s}$
    \item Calculate the residual vector $\mathbf{r} = (\mathbf{A} - \theta \cdot \mathbf{B}) \cdot \mathbf{u} = \mathbf{u_A} - \theta \cdot\mathbf{u_B}$. If $\norm{\mathbf{r}}$ is less than some predefined threshold, return $(\theta, \mathbf{u})$.
    \item Approximately solve the correction equation for $\mathbf{\delta}$,
    \begin{equation}
        \mathbf{F} \cdot \mathbf{\delta} = - \mathbf{r},
    \end{equation}
    with a chosen preconditioner $\mathbf{F}$. In the Jacobi-Davidson scheme, $\mathbf{F} = (\mathbf{I} - \mathbf{u_B} \cdot \mathbf{u}^\dagger)(\mathbf{A} - \theta \cdot \mathbf{B})(\mathbf{I} - \mathbf{u} \cdot \mathbf{u_B}^\dagger)$
    \item Orthonormalize $\mathbf{\delta}$ with respect to the subspace, the columns of $\mathbf{V}$. This can be done with a Gram-Schmidt procedure:
    \begin{enumerate}
        \item for $\mathbf{v} \in \text{Cols}(\mathbf{V})$ \\ 
        $\delta := \delta - \mathbf{v} \cdot (\mathbf{v}^\dagger \cdot \mathbf{B} \cdot \mathbf{\delta})$
        \item $\mathbf{x_t} = \mathbf{\delta}/\norm{\mathbf{\delta}}$, where $\norm{\mathbf{\delta}} = \sqrt{\mathbf{\delta}^\dagger \cdot \mathbf{B} \cdot \mathbf{\delta}}$
    \end{enumerate}
    \item Append $\mathbf{x_t}$ to the subspace. $\mathbf{V} := [\mathbf{V}, \mathbf{x_t}]$, $\mathbf{V_A} := [\mathbf{V_A},\mathbf{A}\mathbf{x_t}]$, and $\mathbf{V_B} := [\mathbf{V_B},\mathbf{B}\mathbf{x_t}]$
    \item Repeat steps 2-8 until a desired convergence for $\norm{\mathbf{r}}$ has been achieved.
\end{enumerate}
To ensure the size of the subspace doesn't grow too large, restarts should be taken advantage of. This can be accomplished very easily by taking a number of the eigenvectors in the subspace problem that are closest to the desired target value as the new subspace. In detail, if $|\text{Cols}(\mathbf{V})|$ is greater than or equal to some $\mathbf{V_{max}}$, define $\mathbf{N}$ as a matrix whose columns are a $\mathbf{V_{restart}}$ number of eigenvectors of the subspace problem (ie. the eigenproblem with matrices $\mathbf{A^\prime}$, $\mathbf{B^\prime}$) with eigenvalues closest to $t$. The new subspace matrices can then be calculated, $\mathbf{V} := \mathbf{V} \cdot \mathbf{N}$, $\mathbf{V_A} := \mathbf{V_A} \cdot \mathbf{N}$, and $\mathbf{V_B} := \mathbf{V_B} \cdot \mathbf{N}$. We find $\mathbf{V_{max}} = 25$ and $\mathbf{V_{restart}} = 5$ work reasonably well for matrices of any dimension. It is also useful to update the target value $t$, when the residual norm $\norm{\mathbf{r}}$, is smaller than some threshold, and to continue updating it to $\theta$ if the corresponding eigenvector has a smaller residual norm.


\section{\label{sec:3}Computational Details}

Given a wavefunction, a VMC run utilizing the LM optimizer can be performed efficiently if the primitives $E_L[n]$, $\mathbf{g}_i[n]$, and $\mathbf{h}_i[n]$ can be computed with a low computational cost.

For the orbital space algorithm, previous publications have demonstrated that the local energy and $\mathbf{g}_i[n]$ can be computed at a cost of $O(N^2)$ if intermediate values are pre-calculated\cite{sabzevari2018improved,mahajan2019symmetry}. The third term $\mathbf{h}_i[n]$, can be naively calculated at a cost of $O(N^4)$, but by considering the gradient of the local energy with respect to the $i\text{th}$ parameter,
\begin{equation}
    \partial_i E_L[n] = \mathbf{h}_i[n] - E_L[n] \cdot \mathbf{g}_i[n],
\end{equation}
one can calculate the last primitive using,
\begin{equation}
    \mathbf{h}_i[n]  = \partial_i E_L[n] + E_L[n] \cdot \mathbf{g}_i[n].
\end{equation}
Noting that the local energy can be calculated directly from the Jastrow and Hartree-Fock parameters at a cost of $O(N^3)$, a reverse accumulation algorithmic differentiation scheme will calculate the gradient at the same asymptotic cost, albeit with a fairly large pre-factor\cite{linnainmaa1976taylor,griewank2008evaluating}. To do this we turn to the Stan Math library, an excellent C++ implementation of algorithmic differentiation, that was very easily extendable to linear algebra with complex numbers\cite{carpenter2015stan}. 

For the real space algorithm, we comment on the computational costs of calculating each primitive, the details of which will be a part of a forthcoming publication. The local energy can be calculated at a cost of $O(N^2)$, while $\mathbf{g}_i[n]$ and $\mathbf{h}_i[n]$ are calculated at a cost of $O(N^3)$. 
The expressions for these values are very similar to those found in reference \cite{filippi2016simple}, with the added complication that the reference is composed of complex variables and is projected onto the correct symmetry before any derivatives are taken.

For both the real space and orbital space wavefunctions, the VMC algorithm for sampling the energy and the matrices
in the direct linear method scales as $O(N^4)$ when a system size independent stochastic error is required. This scaling comes about by using the fact that the primitives are serially correlated, with a correlation length that can be assumed to increase linearly with the size of the system. Thus by only sampling the expensive primitives every $O(N)$ Monte Carlo moves, and having to perform a total of $O(N^2)$ moves to obtain a system-size independent error, yields an overall scaling of $O(N^4)$. 

To stabilize the linear method we add a diagonal shift to the $\mathbf{\bar{H}}$ matrix, which acts as a trust radius to the eigenvector solution. This diagonal shift decays by a multiplicative factor every iteration of the linear method until it reaches a value of $10^{-6}$.  We find that an initial shift of $0.1$ and a decay factor of $0.65$ yield excellent results, and will be used for the majority of the calculations in this paper. Furthermore, to guarantee a large step in a poor update direction is never taken, we employ a short correlated sampling run with wavefunctions created from various step sizes multiplied by the LM update. The step size that generated the lowest relative energy is taken as the wavefunction to begin the next VMC iteration. In detail, we calculate new wavefunctions with the original parameters plus $0.01, 0.05, 0.1, 0.5, 1.0$ times the LM update, 
and perform a VMC run to calculate the energy using the median value step size, and then estimate the energies for the other wavefunctions with,
\begin{equation}
    E = \frac{\expval{E_L[n] \cdot \frac{\abs{\bra{n}\ket{\Psi^\prime}}^2}{\abs{\bra{n}\ket{\Psi}}^2}}_{\rho_n}}{\expval{\frac{\abs{\bra{n}\ket{\Psi^\prime}}^2}{\abs{\bra{n}\ket{\Psi}}^2}}_{\rho_n}},
\end{equation}
where $\ket{\Psi}$ is the wavefunction that the random walk is performed with, and $\ket{\Psi^\prime}$ is a ``slave'' wavefunction that should not be too different from $\ket{\Psi}$. This expression minimizes the relative error between the averaged energies, and allows a short VMC run to be used to calculate energies that can then be compared between each other with high accuracy. We use $0.35$ times $n_{\text{s}}$ number of iterations to perform correlated sampling.

When optimizing a large system with the LM, the linear expansion of the wavefunction can be a poor representation of the working non-linear ansatz, particularly far from the global minumum. 
When this occurs, none of the strategies discussed above will result in decent convergence. To remedy this, we turn to the first order SGD method, AMSGrad, at the beginning of the optimization. This optimizer, developed in the machine learning community, has given us comparable convergence to SR in every practical example our group has attempted, in addition to requiring significantly less overhead. Although the algorithm has been discussed in a previous publication, here we briefly restate the AMSGrad method, since it is a critical part of the presented algorithm and remedies a known problem with the LM\cite{corrsys}.
A general algorithm for adaptive SGD methods is the following,
\begin{equation}
\mathbf{m}^{(i)} = f(\mathbf{G}^{(0)},\cdots, \mathbf{G}^{(i)}),
\end{equation}
\begin{equation}
\mathbf{n}^{(i)} = g(\mathbf{G}^{(0)},\cdots, \mathbf{G}^{(i)}),
\end{equation}
\begin{equation}
\delta p_j = -\alpha \cdot \mathbf{m}_j^{(i)}/\sqrt{\mathbf{n}^{(i)}_j},
\end{equation}
where $\alpha$ is the stepsize and $f$ and $g$ are some functions that take in all the past gradients and generate vectors $\mathbf{m}^{(i)}$ and $\mathbf{n}^{(i)}$. These are then used to calculate the updates $\{\delta p_i\}$. Adaptive SGD methods differ in the chosen functions, $f$ and $g$. In AMSGrad, $f$ and $g$ respectively calculate the exponentially decaying moving average of the first and second moment of the gradients $\{\mathbf{G}^{(i)}\}$,
\begin{equation}\label{eq:m1}
\mathbf{m}^{(i)}_j = (1-\beta_1) \cdot \mathbf{m}^{(i-1)}_j + \beta_1 \cdot \mathbf{G}^{(i)}_j
\end{equation}
\begin{equation}\label{eq:m2}
\mathbf{n}^{(i)}_j = \max(\mathbf{n}^{(i-1)}_j, (1-\beta_2) \mathbf{n}^{(i-1)}_j +  \beta_2(\mathbf{G}^{(i)}_j \cdot \mathbf{G}^{(i)}_j),
\end{equation}
with the caveat that the second moment at iteration $(i)$ is always greater than at iteration $(i-1)$, i.e. $\mathbf{n}^{(i)}_j \geq \mathbf{n}^{(i-1)}_j$. This ensures that the learning rate, $\alpha/\sqrt{\mathbf{n}^{(i)}_j}$, is a monotonically decreasing function of the number of iterations. This was shown to improve the convergence of AMSGrad for a synthetic problem over other adaptive SGD methods\cite{reddi2019convergence}. When optimizing with AMSGrad, unless otherwise specified, we have used the aggressive parameters, $\alpha = 0.01, \beta_1 = 0.1, \beta_2=0.01$. When starting off an aLM optimization with AMSGrad we use the more stable parameters, $\alpha = 0.001, \beta_1 = 0.1, \beta_2=0.001$.


\section{\label{sec:4}Results and Discussion}

We begin by presenting results on a couple of atomic systems with real space Jastrow wavefunctions, and then discuss results for various realistic and model molecular systems with orbital space Jastrow wavefunctions. The focus of the analysis will be on comparing aLM to LM and the first order optimization methods, AMSGrad and SR. Integrals and initial GHF parameters for all calculations were generated by Pyscf\cite{sun2018pyscf}.

\subsection{Beryllium Atom - Real Space}

Table \ref{tab:rBe} shows VMC results for the Beryllium atom using various optimizers. To investigate the direct LM implementation, no AMSGrad iterations were used in the aLM optimization. This is a small system consisting of 4 electrons in 17 orbitals yielding 364 total variables, yet it is already evident that the aLM outperforms the original LM, as well as the cheaper but less robust SGD method, AMSGrad.

\begin{table}[h!]
    \centering
    \begin{tabular}{ccccc}
        \hline
        \hline
        Optimizer & Energy & Iteration & Wall Time & Time/Iteration \\
        \hline
        LM & $-14.667$ & $21$ & $123.84$ & $5.90$ \\
        aLM & $-14.667$ & $8$ & $23.52$ & $2.94$ \\
        AMSGrad & $-14.666$ & $181$ & $128.61$ & $0.71$ \\ 
        \hline
    \end{tabular}
    \caption{Real space Beryllium atom optimization using a J-$KS_z$GHF ansatz, in a Slater TZP all electron basis\cite{van2003optimized}. Energies are in units of Hartrees, and time is in seconds. The same number of samples were used for each macroiteration to ensure an error of $\sim0.5$ mHa was reached for all calculations. Aggressive AMSGrad parameters were used consisting of $\alpha = 0.01, \beta_1 = 0.1, \beta_2=0.01$. As a comparison, the energy from exact diagonalization in an aug-cc-pVQZ basis is $-14.640$ Ha.}
    \label{tab:rBe}
\end{table}

\subsection{Neon Atom - Real Space}

\begin{figure}[h!]
    \centering
    \resizebox{\linewidth}{!}{\input{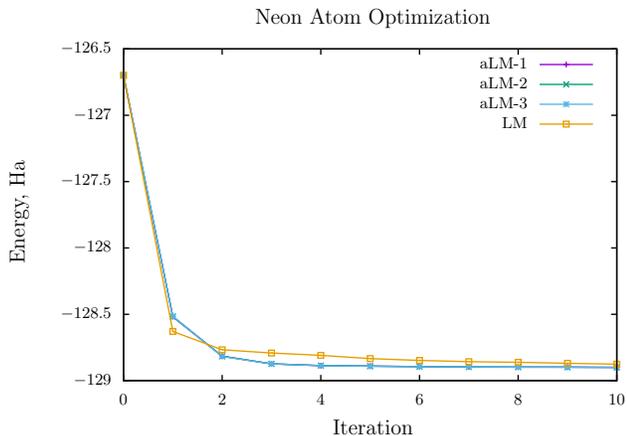}}
    \caption{Optimization curve for real space Neon atom. Curves correspond to the entries of Table \ref{tab:rNe}, where aLM-1 corresponds to the loosest tolerances for aLM, and aLM-3 corresponds to the tightest.}
    \label{fig:rNe}
\end{figure}

Figure \ref{fig:rNe} and Table \ref{tab:rNe} illustrate optimization of the Neon atom. This is a system of 10 electrons in 10 orbitals giving a wavefunction that consists of 492 total variables. We compare aLM curves with varying tolerances on the Davidson and conjugate gradient solvers, to the usual LM. To investigate the direct LM implementation, no AMSGrad iterations were used in the aLM optimization.

\begin{table}[h!]
    \centering
    \begin{tabular}{ccccc}
        \hline
        \hline
        Optimizer & Energy & dTol & cgTol & Time/Iteration \\
        \hline
        aLM & $-128.907$ & $10^{-3}$ & $10^{-2}$ & $555$ \\
        aLM & $-128.908$ & $10^{-3}$ & $10^{-3}$ & $664$ \\
        aLM & $-128.909$ & $10^{-4}$ & $10^{-3}$ & $761$ \\
        LM & $-128.910$ & - & - & $2489$ \\ 
        \hline
    \end{tabular}
    \caption{Optimized real space Neon atom energies with a J-$KS_z$GHF ansatz, in a Slater DZ all electron basis\cite{van2003optimized}. Energies are in units of Hartrees, and time is in seconds. dTol is shorthand for tolerance of the davidson solver, and likewise for cgTol with respect to the conjugate gradient solver. The same number of samples were used for each macroiteration to ensure an error of $\sim0.5$ mHa was reached for all calculations. As a comparison, the energy from exact diagonalization in an aug-cc-pV5Z basis is $-128.900$ Ha.}
    \label{tab:rNe}
\end{table}

Inspecting Figure \ref{fig:rNe}, we see very little difference in the optimization curves between the different implementations of the linear method, however, the difference in timings and energies in Table \ref{tab:rNe} is more pronounced. This suggests that a schedule of progressively tighter tolerances would ensure fast and robust optimization.

\subsection{Carbon Dimer - Orbital Space}

Here we pivot to orbital space VMC, beginning with the carbon dimer at equilibrium geometry. The bond length is taken to be the experimentally determined value of 1.24244 angstroms\cite{xu2013insights,douay1988new}. This system consists of 12 electrons in 18 orbitals and the resulting wavefunction has a total of 1,530 parameters. Note again, no AMSGrad iterations were used in the aLM optimization to focus on the direct LM implementation.

\begin{figure}[h!]
    \centering
    \resizebox{\linewidth}{!}{\input{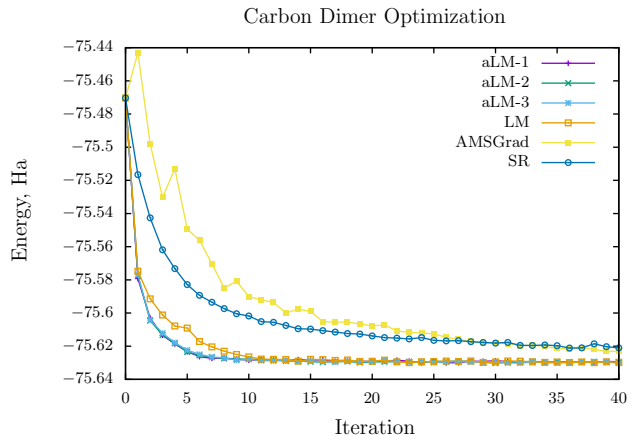}}
    \caption{Optimization curves for orbital space Carbon Dimer with respect to number of iterations. Curves correspond to the entries of Table \ref{tab:C2}, where aLM-1 corresponds to the loosest tolerances for aLM, and aLM-3 corresponds to the tightest.}
    \label{fig:C2}
\end{figure}

\begin{figure}[h!]
    \centering
    \resizebox{\linewidth}{!}{\input{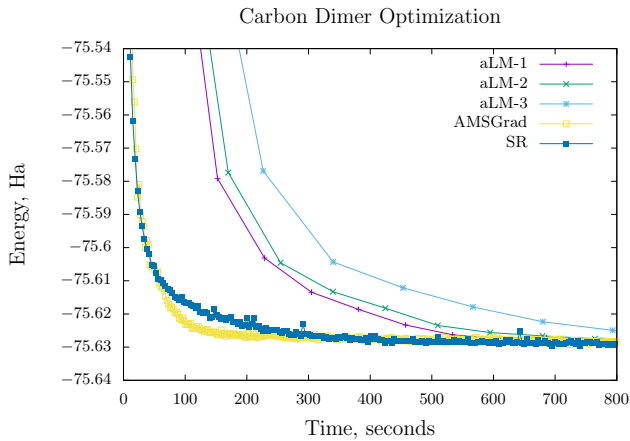}}
    \caption{Optimization curves for orbital space Carbon Dimer with respect to time in seconds. Curves correspond to the entries of Table \ref{tab:C2}, where aLM-1 corresponds to the loosest tolerances for aLM, and aLM-3 corresponds to the tightest. Note that the LM doesn't appear for the first $1000+$ seconds of the plot.}
    \label{fig:C2_time}
\end{figure}

Figure \ref{fig:C2} shows the optimized energy of the Carbon dimer with the number of VMC iterations, while Figure \ref{fig:C2_time} shows the same optimization but with the wall time plotted on the x-axis. It's quite clear from inspecting the two plots that both first order methods (SR and AMSGrad) behave very similarly whether one considers the number of iterations or the total CPU time. The convergence of the various flavors of aLM with different tolerances and LM as a function of the number of iterations is very similar. However, the CPU time for the LM is significantly greater than that of the aLM method and even the first iteration of the LM method takes $1000+$ seconds and does not appear in the plot of Figure~\ref{fig:C2_time}.

\begin{table}[h!]
    \centering
    \begin{tabular}{cccccc}
        \hline
        \hline
        Optimizer & Energy & dTol & cgTol & Iteration & Time/Iteration \\
        \hline
        aLM & $-75.630$ & $10^{-3}$ & $10^{-2}$ & $24$ & $76.2$ \\
        aLM & $-75.630$ & $10^{-3}$ & $10^{-3}$ & $16$ & $84.9$ \\
        aLM & $-75.630$ & $10^{-4}$ & $10^{-3}$ & $17$ & $113.3$ \\
        LM & $-75.630$ & - & - & $23$ & $1485.7$ \\
        AMSGrad & $-75.630$ & - & - & $397$ & $2.5$ \\ 
        SR & $-75.630$ & - & - & $319$ & $3.8$ \\ 
        \hline
    \end{tabular}
    \caption{Optimized orbital space $C_2$ molecule energies with a J-$KS_z$GHF ansatz, in a 6-31G all electron basis. Lowdin's orbital localization scheme was used to define the sampled Hilbert space\cite{lowdin1950non}. Energies are in units of Hartrees, and time is in seconds. dTol is shorthand for tolerance of the davidson solver, and likewise for cgTol with respect to the conjugate gradient solver. Aggressive parameters were used for SR and AMSGrad, with a stepsize of 0.1 and diagonal shift of 0.001 for SR, and $\alpha = 0.01, \beta_1 = 0.1, \beta_2=0.01$ for AMSGrad. The initial energy was $-75.470$ Ha, and the same number of samples were used for each macroiteration to ensure an error of $\sim0.5$ mHa was reached for all calculations. As a comparison, the energy from exact diagonalization in the same basis is $-75.643$ Ha.}
    \label{tab:C2}
\end{table}

Table \ref{tab:C2} shows that all tolerances optimized to the same energy. It should be noted that tighter tolerance does ensure convergence to the final answer in a smaller number of iterations, although this seems to be balanced out by the fact that each iteration of the looser tolerances costs less. A schedule of progressively tightening tolerances again seems like the most efficient course of action
. In Figure \ref{fig:C2}, we also see that the first order methods optimize very well in the first $\sim20$ iterations, but are hindered by long tails, needing $\sim300$ more iterations to converge to the final value. This motivates the use of a first order method in the initial optimization of large systems, as the time per iteration of these methods is at least an order of magnitude cheaper than any of the variants of LM. We also remark on the fact that SR and AMSGrad give very comparable optimization curves for this system, again noting the significantly smaller memory overhead required by AMSGrad. AMSGrad amounts to a single loop over the parameters of the wavefunction, while SR requires storing the primitives for the overlap matrix, and then solving a system of linear equations. While this extra asymptotic cost for SR is comparable to the proposed scheme for aLM, aLM's unmatched optimization make this cost worth the effort.

\subsection{$\mathbf{H}_{50}$ - Orbital Space}

Table \ref{tab:H50} illustrates the optimization of a chain of 50 Hydrogen atoms at a bond length of 2.0 Bohr (1.058354 angstroms) using various optimizers. This is a system of 50 electrons in 50 orbitals and results in an ansatz with 15,050 parameters.

\begin{table}[h!]
    \centering
    \begin{tabular}{cccc}
        \hline
        \hline
        Optimizer & Energy & Iteration & Avg. Time per Iteration \\
        \hline
        SR & $-26.886$ & $247$ & $108.3$ \\
        \hline
        AMSGrad & $-26.886$ & $289$ & $104.7$ \\
        AMSGrad & $-26.905$ & $1146$ & $104.5$ \\
        AMSGrad & $-26.908$ & $1898$ & $104.6$ \\
        \hline
        aLM* & $-26.713$ & $25$ & $109.4$ \\
        aLM & $-26.886$ & $36$ & $335.4$ \\
        aLM & $-26.908$ & $69$ & $622.2$ \\
        aLM & $-26.910$ & $89$ & $796.9$ \\
        \hline
    \end{tabular}
    \caption{$H_{50}$ chain optimization using the orbital space J-$KS_z$GHF ansatz, in a STO-6G basis. Lowdin's orbital localization scheme was used to obtain the localized orbitals for the Jastrow factor\cite{lowdin1950non}. Energies are in units of Hartrees, and time is in seconds. The same number of samples were used for each macroiteration to ensure an error of $\sim0.5$ mHa was reached for all calculations. Aggressive parameters were used for SR and AMSGrad, with a stepsize of 0.1 and diagonal shift of 0.001 for SR, and $\alpha = 0.01, \beta_1 = 0.1, \beta_2=0.01$ for AMSGrad. For aLM the loosest tolerances corresponding to $\text{dTol} = 10^{-3}$, $\text{cgTol} = 10^{-2}$ were used. 25 AMSGrad iterations were done to start off the aLM run, using the less aggressive parameters (this is what results in the differing Avg. Time per Iteration values for the aLM), the energy at which LM iterations start is tabulated with an asterisk.}
    \label{tab:H50}
\end{table}

The lowest energy for each optimizer is the converged value for that optimizer. 
SR and AMSGrad optimize to $-26.886$ Ha in a similar number of iterations, and have very comparable timings per iteration. However, AMSGrad continues to optimize to an energy more than 20 mHa lower than SR, but it's quite clear there is a very long tail as it takes more than 700 more iterations to lower the energy by another 3 mHa. As expected, aLM optimizes significantly faster reaching the SR energy in 36 iterations (25 AMSGrad with the less aggressive parameters plus 16 iterations with aLM and the loose tolerances), then reaching the AMSGrad energy in 33 more iterations, and reaching an energy lower than either of the other optimizers in 89 total iterations. Remarkably, when comparing total time to reach a common converged energy, the aLM is cheaper than SR and AMSGrad. We expect this to be further improved if hand-coded analytic expressions for all the primitives can be implemented as the algorithmic differentiation scheme carries a large computational prefactor.

\subsection{Hydrogen chains - Orbital Space}

In this section, we investigate practical scaling of the aLM algorithm, considering only the time spent to solve the generalized eigenvalue problem. Calculations were performed on linear hydrogen chains consisting of 10 to 60 atoms at a bond length of 2.0 Bohr in a STO-6G basis.

\begin{figure}[h!]
    \centering
    \resizebox{\linewidth}{!}{\input{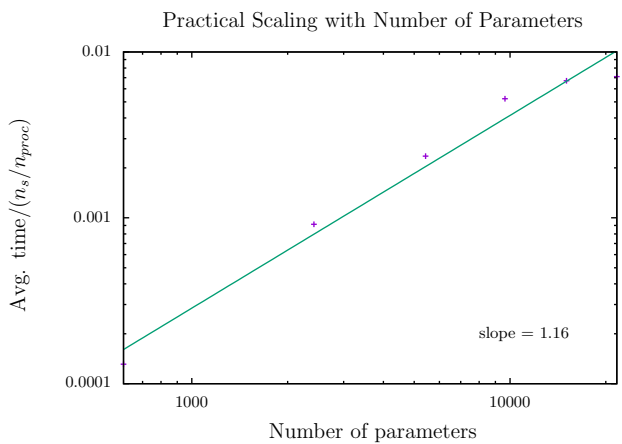}}
    \caption{Average wall time of Jacobi-Davidson solver per Monte Carlo sample per process vs. number of parameters plotted on a log-log scale. Loose tolerances of $\text{dTol} = 10^{-3}$, $\text{cgTol} = 10^{-2}$ were used for all calculations. This suggests the asymptotic scaling of the solver is $O(n_{\text{s}}(n_{\text{var}})^{1.16}/n_{\text{proc}})$.}
    \label{fig:params_scaling}
\end{figure}


The computational experiments suggest that the solver has an asymptotic cost that is slightly higher than the application of the matrix on a vector, a very satisfactory result.
Inserting $n_{\text{var}} = O(N^2)$, we attain an $O(n_{\text{s}} N^{2.32} /n_{\text{proc}})$ cost as a function of system size.

\subsection{98-site Hubbard model - Orbital Space}

Here we discuss the 98 site Hubbard model with $U/t = 2$. A system of 98 electrons in 98 orbitals, this ansatz results in 57,722 parameters. We use this example to investigate any bias in the aLM optimizer. Due to the fact the matrix elements of $\mathbf{\bar{H}}$ and $\mathbf{\bar{S}}$ have stochastic noise, any eigenvector will inherit a non-linear bias, which is expected to increase with the size of the matrix. Checkpoints in the optimization can be seen in Table \ref{tab:98-hubbard}. The converged energy per electron using the aLM optimizer is -1.19208 $t$ and Green's function Monte Carlo (GFMC) reference energy is -1.19622 $t$\cite{leblanc2015solutions}. Error on the energy per electron values are 0.00001 $t$.

\begin{figure}[h!]
    \centering
    \resizebox{\linewidth}{!}{\input{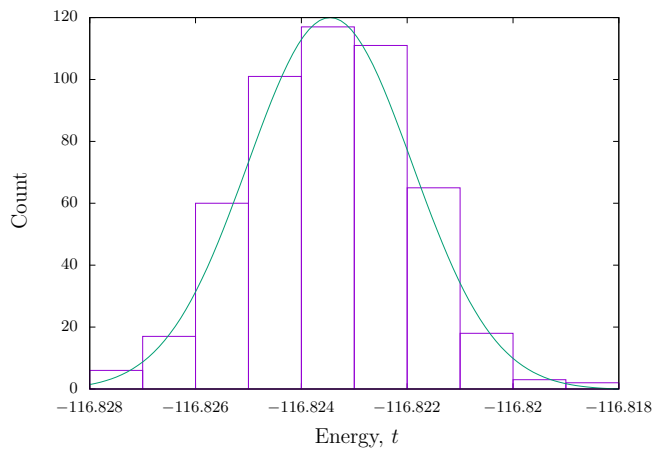}}
    \caption{Distribution of energies from performing 500 SGD iterations on the optimized aLM wavefunction. A small stepsize of 0.0001 was used for SGD.}
    \label{fig:Hubbard98}
\end{figure}

With the converged wavefunction, we perform 500 iterations using plain SGD with a small step size of 0.0001. SGD is an unbiased optimizer and if the aLM results in any bias it should be evident in the distribution of energies. In Table \ref{fig:Hubbard98}, we see that the distribution of energies result in a Gaussian, and can be presumed to be from noise in the gradient as the wavefunction is bounced around the minimum. The converged aLM energy falls directly in the middle bin of the histogram.

\begin{table}[h!]
    \centering
    \begin{tabular}{cc}
        \hline
        \hline
        Iteration & Energy  \\
        \hline
        $0$ & $-113.032$ \\
        $75$ & $-116.258$ \\
        $93$ & $-116.824$ \\
        \hline
    \end{tabular}
    \caption{98-site Hubbard model optimization using a J-$KS_z$GHF ansatz. Energies are in units of $t$, and time is in seconds. Errors on energies are $\sim0.001$ $t$. The loosest tolerances corresponding to $\text{dTol} = 10^{-3}$, $\text{cgTol} = 10^{-2}$ were used for aLM. 75 AMSGrad iterations were done to start off the aLM run, using the less aggressive parameters.}
    \label{tab:98-hubbard}
\end{table}

\subsection{Iron(II) Porphyrin - Orbital Space}

We apply the orbital space J-$KS_z$GHF wavefunction and the aLM optimizer to calculate the energy of the Fe(II) porphyrin (Fe(P)) cluster in the 32 electrons, 29 orbitals active space of Li Manni\cite{li2016combining}. The main aim of the current calculation is to investigate whether VMC calculates the same energy gap between the ${}^5\text{A}_{1g}$ and ${}^3\text{B}_{1g}$ states as the near exact Semi-stochastic Heat-bath Configuration Interaction (SHCI) algorithm\cite{sharma2017semistochastic,holmes2016heat,smith2017cheap}.

\begingroup
\setlength{\tabcolsep}{5pt}
\begin{table}[h!]
    \centering
    \begin{tabular}{cccc}
        \hline
        \hline
        Method & ${}^5\text{A}_{1g}$ & ${}^3\text{B}_{1g}$ & $\Delta({}^5\text{A}_{1g} - {}^3\text{B}_{1g})$ \\
        \hline
        RHF & -2244.753 & -2244.394 & -0.359 \\
        VMC  & -2245.025 & -2244.998 & -0.027 \\
        GFMC & -2245.029 & -2245.002 & -0.027 \\
        Exact(SHCI) & -2245.029 & -2245.002 & -0.027 \\
       \hline
    \end{tabular}
    \caption{Fe(II) porphyrin energy results in Hartrees for the ${}^5\text{A}_{1g}$ and ${}^3\text{B}_{1g}$ states in a cc-pVDZ basis. RHF is restricted open-shell Hartree Fock. A HCI-SCF calculation was performed in the active space with a 
    tight tolerance to get a highly accurate variational result. Semi-stochastic perturbation theory was performed on top of the HCI wavefunction to obtain a nearly exact, albeit non-variational, result in the active space. The active space orbitals from the HCI-SCF calculation were localized in an intrinsic bonding orbital scheme\cite{knizia2013intrinsic}, and then used as the basis for the VMC and GFMC calculations with a J-$KS_z$GHF ansatz. Sufficient samples were taken to obtain an error of less than 0.5 mHa.}
    \label{tab:Fe(P)}
\end{table}
\endgroup

Table \ref{tab:Fe(P)} presents results on the Fe(P) system, for various \textit{ab-initio} methods. The predicted energies of the VMC calculation are quite good, 
the small discrepancies can be attributed to stochastic noise, and lack of dynamical correlation in the ansatz. GFMC\cite{runge1992quantum,trivedi1990ground,sorella1998green,van1994fixed} on top of the optimized J-$KS_z$GHF wavefunction, give results equivalent to the SHCI energies.
The gap predicted by either of the Monte Carlo calculations agrees with the SHCI calculations up to 1 mHa. It is quite remarkable that a wavefunction with merely 5,423 parameters gives results comparable to a CI wavefunction with more than 5 million parameters, a true testament to the ability of these non-linear ansatzes to capture electron correlation. We want to note however, the Jastrow factors of the ansatz are heavily dependent on the orbitals that they are defined in. We localized active space orbitals from an HCI-SCF calculation, and differing localization schemes would result in differing energies, and it is not clear how to choose a basis to achieve the correct answer irrespective of the system, the orbitals, and the localization method. Alternatively, Neuscamman \textit{et al.} have demonstrated that one can optimize the local orbitals simultaneously with the other variational parameters, and further improve the effectiveness of the Jastrow factor\cite{neuscamman2016improved}. We also note that the correlation length for sampling the optimized wavefunctions is $\sim 1.5$; almost every move yields an uncorrelated sample. We believe this is due to the efficient moves generated by the continuous time Monte Carlo algorithm.

\section{\label{sec:5}Conclusion}

In this article, we have detailed an efficient algorithm of the linear method to optimize non-linear wavefunction ansatzes that exhibit strong coupling between their parameters. We have demonstrated that the optimizer functions equally well for wavefunctions with real or orbital space Jastrow factors and results in an order of magnitude reduction in wall time compared to the naive implementation.

For wavefunctions with real space Jastrow factors, our work here indicates that the aLM is the optimizer of choice, as the first order methods struggle to optimize consistently. During the writing of this article we became aware of a related study by the group of Neuscamman \textit{et al.} comparing various optimizers\cite{otis2019complementary}. They suggest that the most efficient convergence is obtained when the bulk of the iterations are carried out using an adaptive stochastic gradient descent interspersed with the linear method (in their work they use the BLM). This conclusion is not in disagreement with ours here, as both studies argue that 
exclusively using a gradient descent algorithm often results in unmanageably slow and poor convergence.

For wavefunctions with orbital space Jastrow factors, the aLM optimizer brings the computer time required to use the linear method to the same order of magnitude as that of the first order methods: AMSGrad and SR. Our results do suggest that for challenging optimization problems the linear method will more consistently optimize to a lower energy. However, the low cost per iteration of the SGD methods, like AMSGrad, suggests that using AMSGrad away from the minimum and switching over to the aLM closer to the minimum will lead to convergence with the least computational cost. We do expect a further reduction in cost of the aLM if an analytic expression for the local energy gradient can be implemented, as the majority of the computational time in the orbital space algorithm is spent performing algorithmic differentiation, which scales well but comes with the largest prefactor in the algorithm. Future work will focuses on implementing hand-coded analytic expressions for the gradient of the local energy, to bring this prefactor down making the aLM algorithm more computationally favorable relative to the first order methods.

The good agreement between VMC results and the near exact SHCI calculations for the calculated triplet-quintet splitting of the Fe(P) molecule illustrates that the J-$KS_z$GHF ansatz is flexible enough to be effectively used as an active space solver for challenging strongly correlated chemical systems. The polynomial scaling of the VMC algorithm opens up the possibility of using the algorithm to tackle all valence active spaces that can't be treated with methods such as full configuration interaction quantum Monte Carlo, selected configuration interaction, or density matrix renormalization group, all of which scale exponentially with the size of the active space. 

\section{\label{sec:6}Acknowledgements}
The funding for this project was provided by the National
Science Foundation through the grant CHE-1800584. IS was also partially supported by the GAANN program through the DOEd. 

%

\end{document}